\documentclass[fleqn,usenatbib]{mnras}

\usepackage[T1]{fontenc}

\DeclareRobustCommand{\VAN}[3]{#2}
\let\VANthebibliography\thebibliography
\def\thebibliography{\DeclareRobustCommand{\VAN}[3]{##3}\VANthebibliography}

\usepackage{graphicx}	% Including figure files
\usepackage{amsmath}	% Advanced maths commands
\usepackage{amssymb}	% Extra maths symbols
\usepackage{color}
\usepackage{booktabs}
\usepackage{breakcites}
\usepackage[table]{xcolor}
\usepackage{empheq}
\usepackage{newtxtext,newtxmath}

\mathchardef\mhyphen="2D

\newcommand{\di}{\mathrm{d}}

\newcommand{\vlos}{{v}_{\rm los}}

\newcommand{\bfr}{\mathbf{r}}
\newcommand{\bfv}{\mathbf{v}}

\newcommand{\pc}{\,{\rm pc}}
\newcommand{\kpc}{\,{\rm kpc}}
\newcommand{\Myr}{\,{\rm Myr}}
\newcommand{\Gyr}{\,{\rm Gyr}}
\newcommand{\kms}{\,{\rm km\, s^{-1}}}

\newcommand{\hatn}{\hat{\textbf{n}}}
\newcommand{\hats}{\hat{\textbf{s}}}
\newcommand{\hatx}{\hat{\textbf{x}}}
\newcommand{\haty}{\hat{\textbf{y}}}
\newcommand{\hatz}{\hat{\textbf{z}}}

\newcommand{\Msun}{\, \rm M_\odot}
\newcommand{\Msunyr}{\, \rm M_\odot\, yr^{-1}}

\newcommand{\Omegap}{\Omega_{\rm p}}

\usepackage{xcolor}
\definecolor{mypink}{rgb}{0.458, 0.188, 0.478}
\definecolor{myblue}{rgb}{0.0, 0.3, 0.8}

\title[Fuelling the nuclear ring of NGC~1097]{Fuelling the nuclear ring of NGC~1097}

\newcommand{\aifa}{Argelander-Institut f\"{u}r Astronomie, Universit\"{a}t Bonn, Auf dem H\"{u}gel 71, 53121, Bonn, Germany}
\newcommand{\ita}{Universit\"{a}t Heidelberg, Zentrum f\"{u}r Astronomie, Institut f\"{u}r theoretische Astrophysik, Albert-Ueberle-Str. 2, 69120 Heidelberg, Germany}
\newcommand{\eso}{European Southern Observatory, Karl-Schwarzschild-Stra{\ss}e 2, 85748 Garching, Germany}
\newcommand{\mcmaster}{Department of Physics and Astronomy, McMaster University, 1280 Main Street West, Hamilton, ON L8S 4M1, Canada}
\newcommand{\cita}{Canadian Institute for Theoretical Astrophysics (CITA), University of Toronto, 60 St George Street, Toronto, ON M5S 3H8, Canada}
\newcommand{\lyon}{Univ Lyon, Univ Lyon1, ENS de Lyon, CNRS, Centre de Recherche Astrophysique de Lyon UMR5574, F-69230 Saint-Genis-Laval France}
\newcommand{\OSU}{Department of Astronomy, The Ohio State University, 140 West 18th Avenue, Columbus, Ohio 43210, USA}
\newcommand{\CCAPP}{Center for Cosmology and Astroparticle Physics, 191 West Woodruff Avenue, Columbus, OH 43210, USA}
\newcommand{\ljmu}{Astrophysics Research Institute, Liverpool John Moores University, 146 Brownlow Hill, Liverpool L3 5RF, UK}
\newcommand{\mpia}{Max-Planck-Institut f\"ur Astronomie, K\"onigstuhl 17, D-69117 Heidelberg, Germany}
\newcommand{\ugent}{Sterrenkundig Observatorium, Universiteit Gent, Krijgslaan 281 S9,
B-9000 Gent, Belgium}
\newcommand{\ign}{Observatorio Astron{\'o}mico Nacional (IGN), C/Alfonso XII, 3, E-
28014 Madrid, Spain}
\newcommand{\oxford}{Sub-department of Astrophysics, Department of Physics, University of Oxford, Keble Road, Oxford OX1 3RH, UK}
\newcommand{\durham}{Institute for Computational Cosmology, Department of Physics, Durham University, South Road, Durham, DH1 3LE, UK}
\newcommand{\aapf}{NSF Astronomy and Astrophysics Postdoctoral Fellow}
\newcommand{\arizona}{Steward Observatory, University of Arizona, Tucson, AZ 85721, USA}
\newcommand{\anu}{Research School of Astronomy and Astrophysics, Australian National University, Canberra, ACT 2611, Australia}  
\newcommand{\arc}{ARC Centre of Excellence for All Sky Astrophysics in 3 Dimensions (ASTRO 3D), Australia}
\newcommand{\iwr}{Universit\"{a}t Heidelberg, Interdisziplin\"{a}res Zentrum f\"{u}r Wissenschaftliches Rechnen, Im Neuenheimer Feld 205, D-69120 Heidelberg, Germany}
\newcommand{\epfl}{Institute of Physics, Laboratory for galaxy evolution and spectral modelling, EPFL, Observatoire de Sauverny, Chemin Pegais 51, 1290 Versoix, Switzerland}
 \newcommand{\ualberta}{Dept. of Physics, 4-183 CCIS, University of Alberta, Edmonton, Alberta T6G 2E1, Canada}
\newcommand{\tum}{Technical University of Munich, School of Engineering and Design, Department of Aerospace and Geodesy, Chair of Remote Sensing Technology, \\\hspace{2.2mm}Arcisstr. 21, 80333 Munich, Germany}
\newcommand{\cool}{Cosmic Origins Of Life (COOL) Research DAO, coolresearch.io}
\newcommand{\manch}{Jodrell Bank Centre for Astrophysics, Department of Physics and Astronomy, University of Manchester, Oxford Road, Manchester M13 9PL, UK}
\newcommand{\UCSD}{Center for Astrophysics and Space Sciences, Department of Physics, University of California San Diego, 9500 Gilman Drive, La Jolla, CA 92093, USA}
\newcommand{\ari}{Astronomisches Rechen-Institut, Zentrum f{\"u}r Astronomie der Universit{\"a}t Heidelberg, M{\"o}nchhofstra{\ss}e 12-14, 69120 Heidelberg,Germany}

\author[Sormani et al.]{%
Mattia~C.~Sormani,$^{1}$\thanks{E-mail: mattiacarlo.sormani@gmail.com}
Ashley~T.~Barnes,$^{2,3}$
Jiayi~Sun,$^{4,5}$
Sophia~K.~Stuber,$^6$
Eva~Schinnerer,$^6$ \newauthor
Eric~Emsellem,$^{2,7}$ 
Adam~K.~Leroy,$^{8,9}$
Simon~C.O.~Glover,$^1$
Jonathan~D.~Henshaw,$^{10,6}$
Sharon~E.~Meidt,$^{11}$ \newauthor 
Justus Neumann,$^6$ 
Miguel Querejeta,$^{12}$
Thomas~G.~Williams,$^{13,6}$
Frank Bigiel,$^3$
Cosima Eibensteiner,$^3$ \newauthor
Francesca Fragkoudi,$^{14}$ 
Rebecca~C.~Levy,$^{15}$\thanks{\aapf}
Kathryn~Grasha,$^{16,17}$\thanks{ARC DECRA Fellow}
Ralf S.~Klessen,$^{1,18}$ \newauthor
J.~M.~Diederik~Kruijssen,$^{19,20}$ 
Nadine Neumayer,$^6$
Francesca~Pinna,$^6$
Erik W.~Rosolowsky,$^{21}$ \newauthor
Rowan J.~Smith,$^{22}$ 
Yu-Hsuan Teng,$^{23}$
Robin~G.~Tress,$^{24}$
Elizabeth J.~Watkins$^{25}$
\\
$^1$ \ita \\
$^2$ \eso \\ 
$^3$ \aifa \\
$^4$ \mcmaster \\
$^5$ \cita \\
$^6$ \mpia \\
$^7$ \lyon \\
$^8$ \OSU \\
$^9$ \CCAPP \\
$^{10}$ \ljmu \\
$^{11}$ \ugent \\
$^{12}$ \ign \\
$^{13}$ \oxford \\
$^{14}$ \durham \\
$^{15}$ \arizona \\
$^{16}$ \anu \\
$^{17}$ \arc \\
$^{18}$ \iwr \\
$^{19}$ \tum \\
$^{20}$ \cool \\
$^{21}$ \ualberta \\
$^{22}$ \manch \\
$^{23}$ \UCSD \\
$^{24}$ \epfl \\
$^{25}$ \ari
}
\begin{document}
\label{firstpage}
\pagerange{\pageref{firstpage}--\pageref{lastpage}}
\maketitle

\begin{abstract}
Galactic bars can drive cold gas inflows towards the centres of galaxies. The gas transport happens primarily through the so-called bar ``dust lanes'', which connect the galactic disc at kpc scales to the nuclear rings at hundreds of pc scales much like two gigantic galactic rivers. Once in the ring, the gas can fuel star formation activity, galactic outflows, and central supermassive black holes. Measuring the mass inflow rates is therefore important to understanding the mass/energy budget and evolution of galactic nuclei.  In this work, we use CO datacubes from the PHANGS-ALMA survey and a simple geometrical method to measure the bar-driven mass inflow rate onto the nuclear ring of the barred galaxy NGC~1097. The method assumes that the gas velocity in the bar lanes is parallel to the lanes in the frame co-rotating with the bar, and allows one to derive the inflow rates from sufficiently sensitive and resolved position-position-velocity diagrams if the bar pattern speed and galaxy orientations are known. We find an inflow rate of $\dot{M}=(3.0 \pm 2.1)\Msunyr$ averaged over a time span of 40 Myr, which varies by a factor of a few over timescales of $\sim$10 Myr. Most of the inflow appears to be consumed by star formation in the ring which is currently occurring at a rate of ${\rm SFR}\simeq~1.8 \mhyphen 2 \Msunyr$, suggesting that the inflow is causally controlling the star formation rate in the ring as a function of time.
\end{abstract}

\begin{keywords}
galaxies: bar -- galaxies: ISM -- galaxies: kinematics and dynamics -- galaxies: nuclei -- galaxies:individual:NGC~1097
\end{keywords}

%%%%%%%%%%%%%%%%% BODY OF PAPER %%%%%%%%%%%%%%%%%%

\section{Introduction} \label{sec:introduction}

It is well-known that galactic bars can efficiently transport cold gas from galactocentric radii of order $R=$ several kpc down to $R=$ few hundreds pc \citep[e.g.][]{Sellwood1993,Kormendy2004,Garcia-Burillo2005,Garcia-Burillo2009,Holmes2015}. These inflows fuel star formation in nuclear rings \citep{Mazzuca2008,Comeron2010}, galactic outflows \citep{Veilleux2020,Ponti2021}, and central supermassive black holes \citep[e.g.][]{Padovani2017,Combes2021}. Measuring the inflow rates is therefore important to understand what regulates star formation in galactic centres \citep{Kruijssen2014,Armillotta2019,Sormani2020b, Moon2021a,Moon2021b,Henshaw2022}, the formation of nuclear stellar discs \citep{Gadotti2019,Bittner2020,Nogueras-Lara2020b,Sormani2022,deSa-Freitas2022}, and the feeding of active galactic nuclei \citep{Davies2007,Storchi-Bergmann2019}.

\begin{figure}
	\includegraphics[width=\columnwidth]{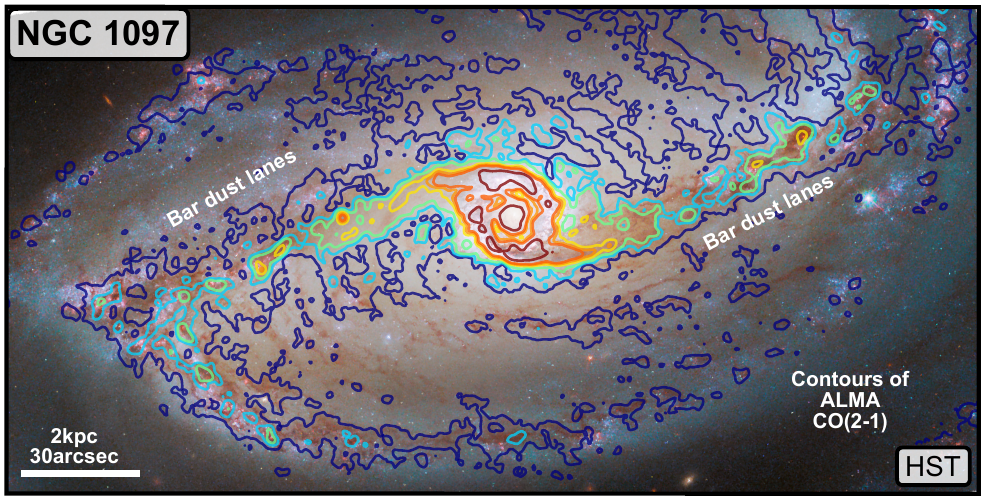}
    \caption{Position of bar dust lanes in the nearby, barred spiral galaxy NGC~1097. We show in the background a (WFC3) {\it HST} image composed of the following filters from projects PID 13413 and 15654 (processed as part of PHANGS-HST; see \citealp{Lee2021}): F275W\,nm (UV) and F336W (u) broad-band in violet, F438W (B) broad-band in dark blue, F547M (Strömgren y) medium-band in cyan, F555W broad-band (V) in green, F814W broad-band (I) in yellow, and F657N (H$\alpha$ + [N II]) in red. Overlaid as coloured contours is the CO(2-1) integrated intensity from the PHANGS-ALMA survey \citep{Leroy2021survey}, in levels of 2, 10, 20, 50, 75, 100, 250\,K\,km\,s$^{-1}$ (increasing from blue to red). Note that the galaxy is rotated in the plane of the sky such that the direction of the kinematic position angle points towards the negative $x$ axis (see Fig.\,\ref{fig:dl}).}
    \label{fig:rgb}
\end{figure}

The transport of gas towards the centre occurs primarily through the two so-called bar ``dust lanes'', one on each side of the bar, which connect the galactic disc at kpc scales to the nuclear rings at few hundred pc scales (see for example Fig.~\ref{fig:rgb}).\footnote{The bar dust lanes are sometimes called ``bar shocks'' because, from a hydrodynamical point of view, they are large-scale shocks  in the interstellar medium. For intuitive explanations of why these shocks form, see for example \citet{Prendergast1983} and \citet{Sellwood1993}.} This has been confirmed by many hydrodynamical simulations of gas flowing in barred potentials \citep[e.g.][]{Athanassoula1992,Englmaier1997,Fux1999,Kim2012,Sormani2015c,Fragkoudi2016,Armillotta2019,Seo2019,Tress2020a}. In particular, these simulations have shown that the velocity vector of the gas in the bar lanes is almost parallel to the lanes in the frame co-rotating with the bar (see for example fig.~3 of \citealt{Athanassoula1992} or fig.~8 in \citealt{Sormani2018c}). The bar lanes therefore act as two gigantic ``galactic rivers'' along which the gas plunges almost radially from the galactic disc at $R=$ several kpc down to the nuclear region at $R=$ few hundred pc. This behaviour can also be clearly seen for example in the movies\footnote{\url{https://www.youtube.com/watch?v=j62sfCTztPg}} of the simulation from \cite{Tress2020a}, where one can follow by eye CO clouds falling along the bar lanes towards the nuclear ring (see link in the footnote). These movies also show that once the CO clouds reach the vicinity of the nuclear ring they are not always accreted immediately onto it: sometimes they do, but sometimes they ``overshoot'', eventually landing on the bar lane on the opposite side, joining its flow to be accreted at a later stage. Observational evidence of this overshooting process was recently obtained by JWST for the barred galaxy NGC 1365 \citep{Whitmore2023}. \citet{Hatchfield2021} further studied the inflow process using Monte Carlo tracer particles to more precisely keep track of the gas flows in a simulation, and quantified the fraction of overshooting gas using an accretion efficiency factor which, for the particular gravitational potential they used, they estimated to be $\epsilon= 30\% \pm 12\%$ when averaged over sufficiently long ($\gtrsim $30 Myr) timescales.

The picture of the inflow process described above suggests that the inflow rates can be estimated directly from observations if we can measure the mass density and flow velocity of cold gas along the bar lanes. Indeed, based on such considerations, \cite{Regan1997} developed a simple geometrical method to estimate the mass inflow rate in the barred galaxy NGC 1530. However, the bar lane width was only marginally resolved in their observations, and they used H$\alpha$ velocities, which trace ionised gas, as a proxy for cold gas velocities since they could not detect CO in the bar lanes due to low sensitivity. \cite{Sormani2019c} adapted the method to derive the bar-driven mass inflow rate in the Milky Way from CO position-position-velocity (PPV) datacubes taking into account the different geometry due to our view through the Galactic plane.

The PHANGS-ALMA survey \citep{Leroy2021survey} opens up new possibilities to derive inflow rates in barred galaxies. This survey has mapped CO${}^{12}$ $J=2\to 1$ line emission in 90 nearby (distance $<$ 20 Mpc) massive star-forming galaxies, the majority of which are barred \citep{Sun2020b,Querejeta2021}. The ALMA CO data has high spatial resolution of 50-150 pc at the distance of the targets and high sensitivity with 1-$\sigma$ noise levels of 0.2–0.3 K per 2.5 $\kms$ channel. It is therefore an ideal dataset to trace the cold gas velocity and mass density in the bar lanes and to measure the mass inflow rates.

In this paper, we build upon the methods of \cite{Regan1997} and \cite{Sormani2019c} and derive the bar-driven mass inflow rate in the barred galaxy NGC~1097 using PHANGS-ALMA CO data. NGC~1097 \citep[e.g.][]{Hummel1987,Barth1995} is a strongly barred galaxy which hosts an intensely star-forming nuclear ring (SFR~$\simeq 1.8\mhyphen 2 \Msunyr$; \citealt{Sandstrom2010,Hsieh2011,Prieto2019,Lopez-Rodriguez2021,Song2021}) with a radius of $\sim$700$\pc$. NGC~1097 is an ideal candidate to determine the inflow rate thanks to its proximity which means high resolution (spatial resolution of $\sim$\,110 pc, see Section~\ref{sec:data}) and for having clearly defined gas-rich bar lanes.

\section{Data} \label{sec:data}

We use ${}^{12}$CO $J=2-1$ data from the PHANGS-ALMA survey (version 4.0; see \citealp{Leroy2021survey} for survey details, and \citealp{Leroy2021pipeline} for in depth discussion of the data reduction, imaging and pipeline). The position-position-velocity (PPV) datacube has a velocity resolution of 2.5 km\,s$^{-1}$ and an angular resolution of 1.7\arcsec, which at the distance of NGC~1097, 13.58\,$\pm$\,2.04\,Mpc \citep{Anand2021}, corresponds to a linear spatial resolution of $\sim$\,110 pc. We assume that NGC~1097 has a position angle of 122.4$^\circ$ and an inclination of 48.6$^\circ$ \citep{Lang2020}. We take $\Omegap=-21.6\, {\rm km\, s^{-1} \, \kpc^{-1}}$ as our fiducial value of the bar pattern speed \citep{Lin2013}. This value is obtained by matching hydrodynamical simulations to the observed morphology of NGC~1097. The negative sign is according to the conventions used in this paper (see Section \ref{sec:deproject}). The properties of NGC~1097 are summarised in Table~\ref{tab:NGC1097}.

    \begin{table}
    	\centering
    	\begin{tabular}{lcc} 
        Property & Value & Reference \\
    		\hline
      Position angle & 122.4$^\circ$ & \citet{Lang2020} \\
      Inclination & 48.6$^\circ$ & \citet{Lang2020} \\
      Distance & 13.58 $\pm$ 2.04 Mpc & \citet{Anand2021} \\
      $\Omegap$ (bar pattern speed) & -21.6$\kms\kpc^{-1}$ & \citet{Lin2013} \\
      $\dot{M}$ (mass inflow rate) & $3.0\pm 2.1 \Msunyr$ & This work \\
      SFR in the nuclear ring & $1.8 \mhyphen 2 \Msunyr$ & See Section~\ref{sec:balance}
    \end{tabular}
            \caption{Properties of NGC~1097.}
    \label{tab:NGC1097}
    \end{table}

\section{Methodology} \label{sec:methodology}

In this section, we describe our methodology to derive the inflow rates. As mentioned in the introduction, this method builds upon the works of \citet{Regan1997} and \citet{Sormani2019c}. The key assumption in all these works (and in the present work) is the same: that the gas velocity vector in the bar lanes is parallel to the lane, and therefore that the bar lanes act as ``galactic rivers'' in which the gas flows. As we shall see below, under this assumption it is possible to derive the \emph{total} gas velocity in the lanes if the galaxy orientation and bar pattern speed are known. The methodology then allows one to calculate the inflow rate if the gas mass density along the bar lanes is also known. As we discuss more in detail below, the largest uncertainty in our derived mass inflow rate comes from the CO-to-H$_2$ mass conversion factor.

This section is structured as follows. In Section~\ref{sec:identification} we describe how we identify the bar lanes in the PHANGS-ALMA CO datacube. In Section~\ref{sec:deproject} we show how to calculate the total gas velocity along the lanes given the line-of-sight component only. In Section~\ref{sec:inflowrate} we explain how we derive the mass density along the lanes and calculate the inflow rates. 

\subsection{Bar lanes identification} \label{sec:identification}

\begin{figure*}
	\includegraphics[width=\textwidth]{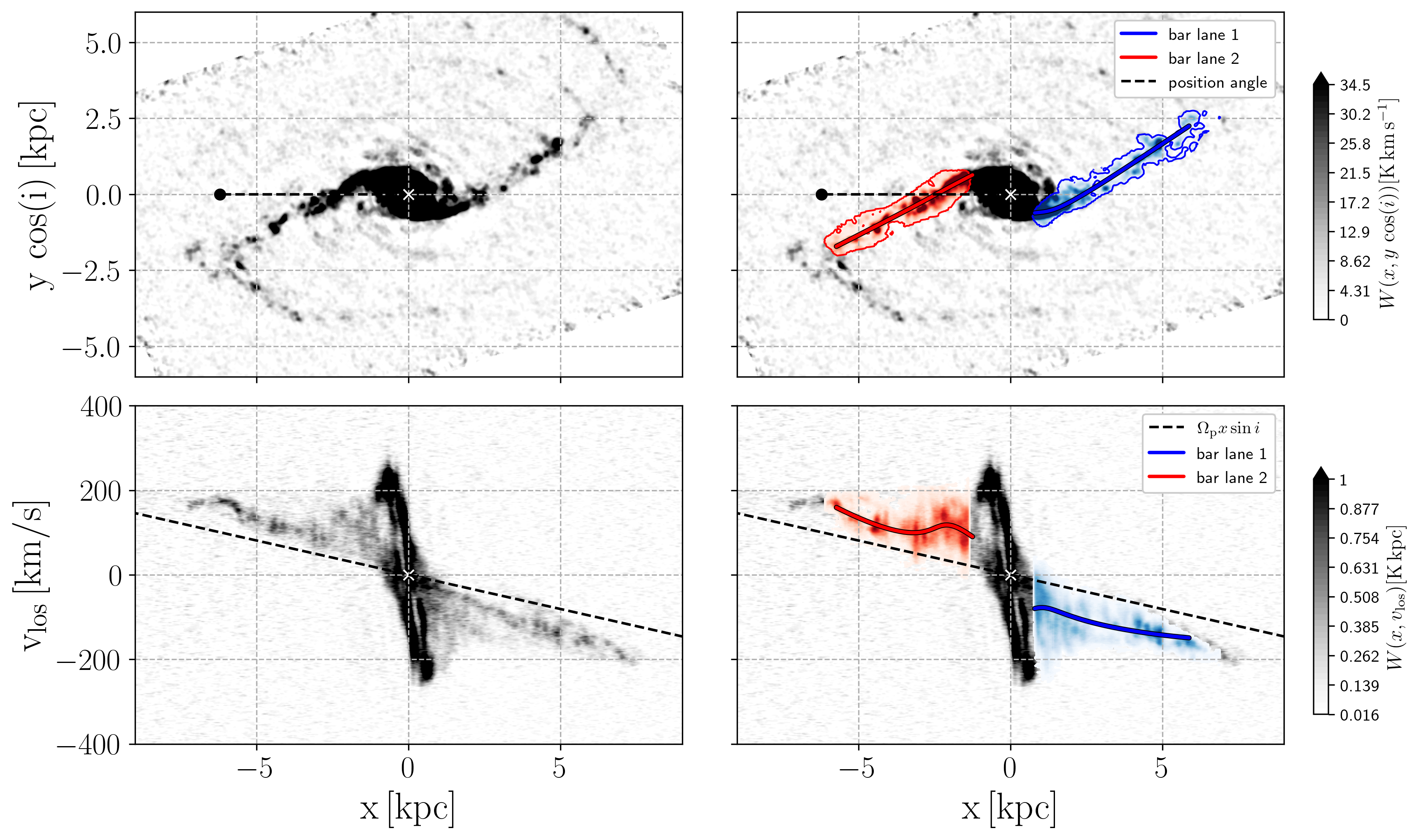}
    \caption{Position-position (top) and position-velocity (bottom) projection of the CO $J=2-1$ datacube of NGC~1097. The galaxy is rotated in the plane of the sky such that the direction of the kinematic position angle points towards the negative $x$ axis (see black dashed line in the top panels). Red and blue contours indicate the emission in the datacube associated with the bar bar lanes. The red and blue lines indicate the spline fits to the bar lanes. The black dashed line in the bottom-right panel indicates the quantity $v_{\rm los}=\Omega_{\rm p} x \sin i$ which appears in the deprojection of the velocity parallel to the bar lanes (see Equation~\ref{eq:vpar}).}
    \label{fig:dl}
\end{figure*}

The first step is visual identification of the bar lanes from the PHANGS-ALMA CO datacube, which are clearly visible in NGC\,1097 (Fig.\,\ref{fig:rgb}). 
The bar lanes are then cut out manually in 3D in position-position-velocity (PPV) space using the {\sc glue}\footnote{https://glueviz.org/} environment, which was chosen for its efficient 3D volume rendering of the emission in PPV space and functionality in separating subsets of a dataset.
 This is done with the following three steps: 
 1) we plot the datacube as a PPV volume rendering; 
 2) we select a data subset by masking in position-position space by tracing a direct path in PPV from the galaxy centre to the bar-end,
 3) we then define a further subset of the position-position space subset by applying an additional mask to the emission that includes the velocity space information. In this mask, we select the emission associated with the bar lanes, which can be clearly seen with a strong velocity gradient towards 0\,km\,s$^{-1}$, and remove any line-of-sight contamination that is clearly distinguishable in velocity space (Fig.\,\ref{fig:rgb} and \ref{eq:vpar}).
Lastly, we omitted the remaining emission within the mask below $\sim$\,$1\sigma$ of the noise - i.e. excluding the insignificant emission. Overall, the procedure is relatively straightforward in the case of NGC\,1097 since there is not much confusion between the bar lanes and unassociated emission in the disc and due to the near face-on orientation, and the strong, bright and relatively isolated bar. Figure~\ref{fig:dl} shows the result for the two bar lanes (show in red and blue) in position-position and position-velocity space.

Once the bar lanes have been identified, we fit a spline through them using the function {\sc splprep} contained in the {\sc scipy.interpolate} python package.\footnote{\url{https://docs.scipy.org/doc/scipy/reference/generated/scipy.interpolate.splprep.html}.} The fitting is done in 3D in the PPV datacube, and each cube point is weighted by the CO brightness of the cube pixel. Note that this  essentially relies on treating PPV as a PPP space (i.e., treating velocity coordinate on the same footing as the spatial coordinate), so the procedure is in principle sensitive to the velocity resolution relative to spatial resolution. In practice, it works well because for our spatial and velocity resolution the number of resolution elements spanned in the velocity direction in a bar lane ($150\kms/2.5\kms\simeq 60$) is roughly comparable to the number of points spanned in the spatial directions ($5\kpc/110\pc\simeq 45$). The result of the spline fitting is shown in the right panels in Figure~\ref{fig:dl} as blue and red curves. In this way, we obtain a spline defined by $N=41$ points, each of them with an associated position and a line-of-sight velocity. 

\subsection{Bar lane deprojection} \label{sec:deproject}

\begin{figure}
	\includegraphics[width=\columnwidth]{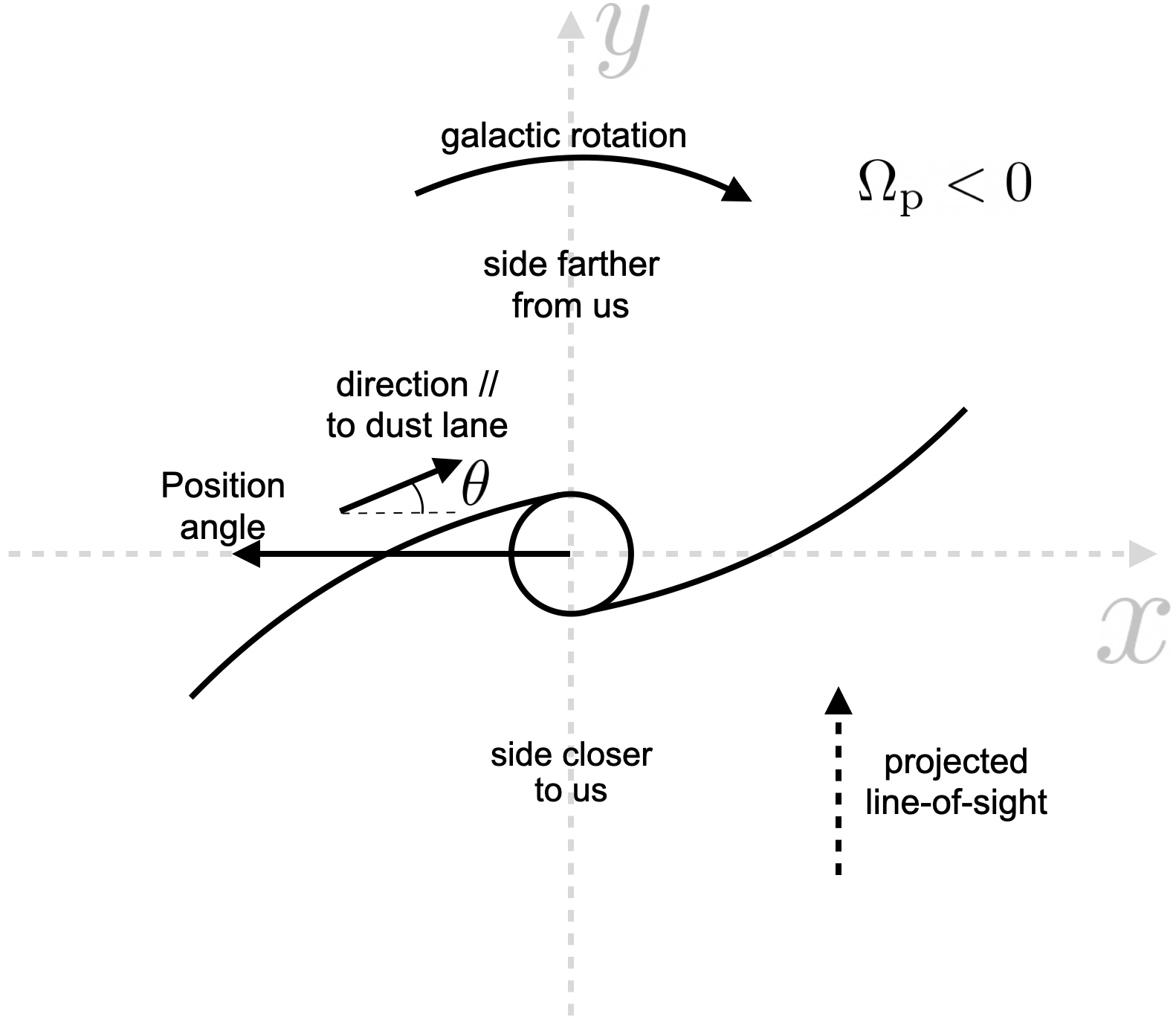}
    \caption{Sketch illustrating the geometry for the bar lane deprojection (see Section~\ref{sec:deproject}). The $z$ axis points out of the page, and it is inclined by an angle $i$ with respect to the line of sight.}
    \label{fig:sketch}
\end{figure}

\begin{figure}
	\includegraphics[width=\columnwidth]{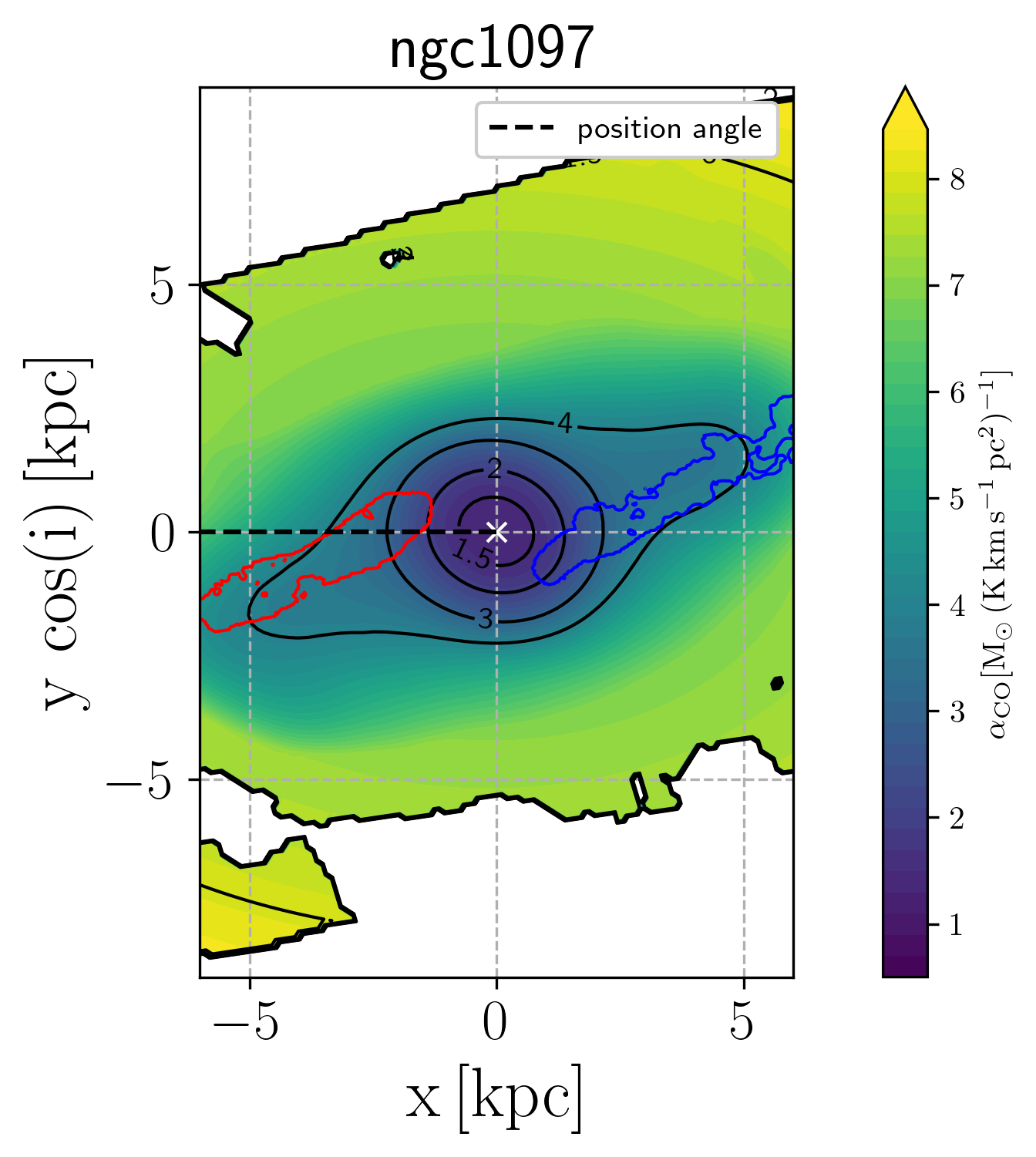}
    \caption{Map of the spatial-dependent $\alpha_{\rm CO}^{(2-1)}$ factor we used to convert from CO intensity to mass. The red and blue contour outline the two bar lanes. The $\alpha_{\rm CO}^{(2-1)}$ varies by a factor of $\sim$2 from one end to the other of each bar lane. The galaxy orientation in the plane of the sky is as in Figs.\ref{fig:rgb} and \ref{fig:dl}.}
    \label{fig:alphaCO}
\end{figure}

Consider a galactocentric Cartesian frame $(x,y,z)$ such that the $z=0$ plane coincides with the midplane of the galaxy (see Fig.~\ref{fig:sketch}). The galaxy rotates around the $z$ axis. We use the convention that the rotation angular velocity is positive if the galaxy rotates anticlockwise, and negative if the galaxy rotates clockwise. NGC~1097 is rotating clockwise\footnote{The direction of rotation is derived assuming that bar lanes are on the leading side of the bar, as it is virtually always the case.} and so the sign of rotation is negative (see sketch in Fig.~\ref{fig:sketch}).

We orient the $(x,y,z)$ frame so that when the galaxy is projected onto the plane of the sky, the direction defined by the galaxy kinematic position angle\footnote{The (kinematic) position angle of a galaxy is defined as the direction of the ``redshifted'' side of the galaxy, measured  relative to the north celestial pole (NCP), turning positive towards East. With the convention used in the figures in this paper, the position angle turns positive counterclockwise.} points towards the negative $x$ direction (see Figs.~\ref{fig:dl} and \ref{fig:sketch}). These conventions have the following implications: (1) the line of nodes coincides with the $x$ axis; (2) for matter in purely circular motion, the side of the galaxy at $x<0$ has positive line-of-sight velocity (it is moving away from us), and the side at $x>0$ has negative line-of-sight velocity (it is moving towards us); (3) deprojecting the galaxy consists in a rotation around the $x$ axis of an angle $0\leq i \leq 90^\circ$ (the inclination angle). Therefore, the $x$ axis remains the same in the projected and deprojected views, while the $y$ axis is stretched by a factor $\cos(i)$. This assumes that all matter in the image is located in the $z=0$ plane; (4) the side of the galaxy at $y<0$ is nearer to (farther from) us if the galaxy rotates clockwise (counterclockwise).

The unit vector in the direction of the line of sight can be written as $\hatn=-\hatz \cos(i) - \haty \sin(i) \eta$, where $\eta=-1$ if the galaxy rotates clockwise and $\eta=+1$ if the galaxy rotates anticlockwise.\footnote{In practice, $\eta$ is the sign of the bar pattern speed.} Let $\bfv(x,y)=v_x \hatx + v_y \haty$ be the velocity in the inertial frame and $\bfv_{\rm bar}(x,y)=v_{{\rm bar},x} \hatx + v_{{\rm bar},y} \haty$ the velocity in the frame rotating at the pattern speed of the bar (we ignore vertical motions). The two are related by
\begin{equation}
    \bfv = \bfv_{\rm bar} + \Omegap \left( \hatz \times \bfr \right)\,,
\end{equation}
where 
\begin{align}
    \Omegap \left( \hatz \times \bfr \right)  = \Omegap \left[ x \haty -y \hatx\right] \,,
\end{align}
and $\Omegap$ is the bar pattern speed, which can take both signs depending on the direction of rotation. The line of sight velocity is then:
\begin{equation} \label{eq:vlos1}
    v_{\rm los} = \bfv \cdot \hatn = - v_{{\rm bar},y} \sin(i) \eta - \Omegap x \sin(i) \eta\,.
\end{equation}
Using the assumption discussed above that in the rotating frame the gas velocity is parallel to the bar lane, we can write $\bfv_{\rm bar}= v_\parallel \hats$ where $\hats = \hatx \cos\theta + \haty \sin\theta$ is the direction parallel to the bar lane and $\theta$ is the angle between the bar lanes and the $x$ axis (see Fig~\ref{fig:sketch}). We take $v_\parallel$ to be positive if the gas moves towards the central ring. Using this, Eq.~\eqref{eq:vlos1} can be rewritten as:
\begin{equation}
    v_{\rm los} = - v_\parallel \sin(\theta) \sin(i)\eta - \Omegap x \sin(i) \eta\,,
\end{equation}
and rearranging we obtain
\begin{equation} \label{eq:vpar}
    v_\parallel = - \frac{v_{\rm los} + \Omegap x  \sin(i) \eta}{\sin(i)\sin(\theta)\eta } \,.
\end{equation}
This equation gives the total gas velocity along the bar lane as a function of observable quantities. The angle $\theta$ is measured from the deprojected spline. $v_{\rm los}$ is the line-of-sight velocity along the spline. $\Omegap$ is the pattern speed of the bar. When $v_\parallel=0$, in our model the gas is not moving at all in the frame of the bar, and from Eq.~\eqref{eq:vpar} we indeed recover the $v_{\rm los}$ that corresponds to rigid body rotation at angular velocity $\Omegap$.

\begin{figure}
	\includegraphics[width=\columnwidth]{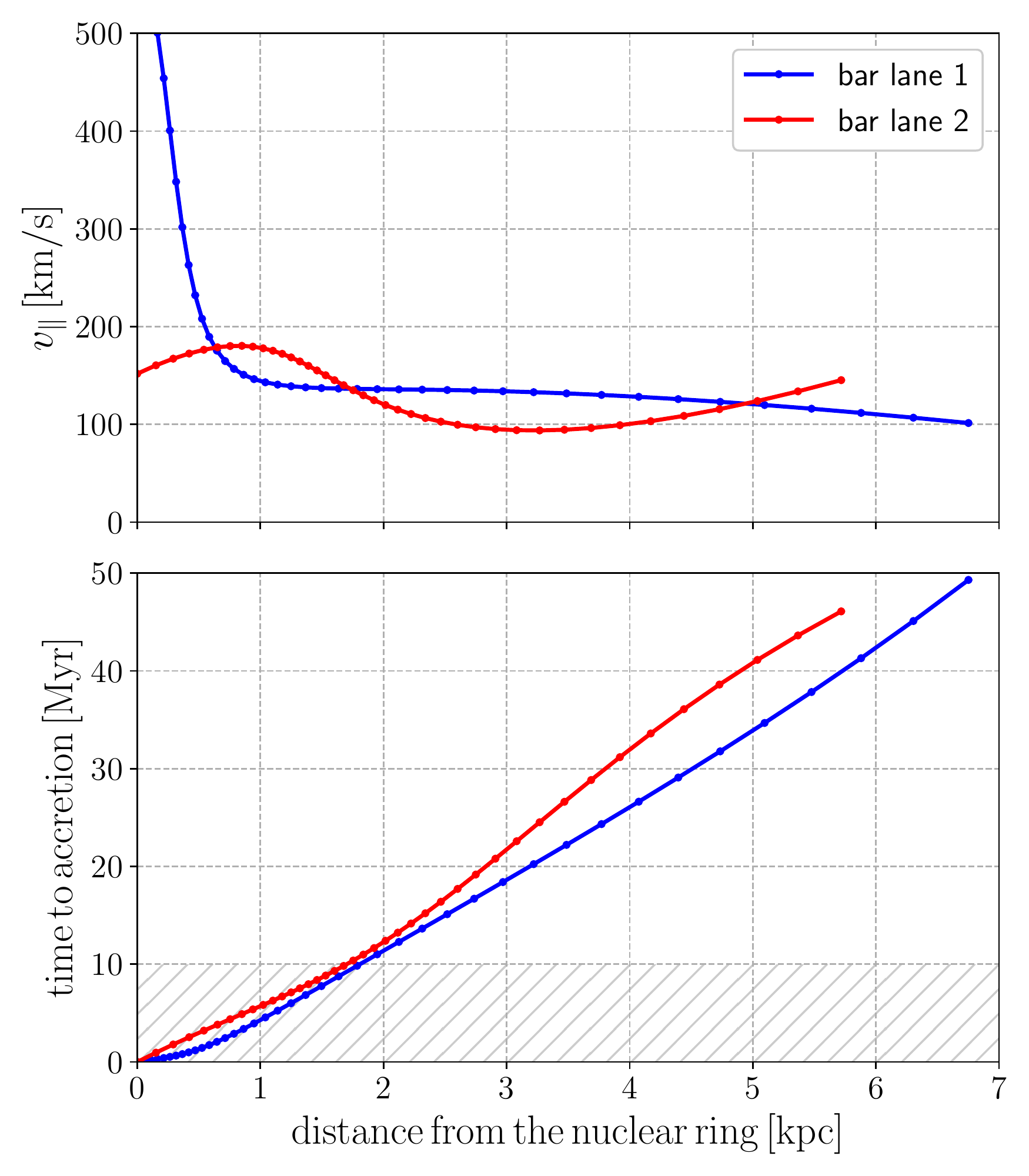}
    \caption{\emph{Top}: total velocity along the bar lanes of NGC~1097 (Eq.~\ref{eq:vpar}), as a function of distance from the nuclear ring calculated in the curved path \emph{along} the bar lane. The sharp increase in the blue curve near the centre in the top panel is likely an artefact due to the bar lane becoming almost parallel to the line of nodes (see Section \ref{sec:results}). \emph{Bottom}: time the gas will take to accrete onto the nuclear ring, calculated using Eq.~\eqref{eq:taccr}. The hatched area shows where the inflow rate is not reliable due to interaction between the bar lanes and the nuclear ring (as in Fig.~\ref{fig:Mdot}).}
    \label{fig:vpar}
\end{figure}

\subsection{Calculation of the inflow rate}
\label{sec:inflowrate}

At this stage we have the following quantities at each point $j$ along the spline:
\begin{itemize}
    \item The deprojected coordinates $(x_j,y_j)$
    \item The velocity $v_{\parallel,j}$ (Equation~\ref{eq:vpar})
\end{itemize}
The spline has $N$ points $j=\{0,\dots,N-1\}$. The point $j=0$ corresponds to the bar lane end closest to the nuclear ring, while $j=N-1$ corresponds to the bar lane end farthest from the centre.

\subsubsection{Time to accretion}

Assuming steady-state, the time it takes for a gas parcel to go from spline point $j$ to point $j-1$ is: 
\begin{equation} 
    \di t_{{\rm accr}, j}  =  \frac{\di s_j }{v_{\parallel, j}} \,,
\end{equation}
where $\di s_j = \sqrt{(x_{j-1}-x_{j})^2 + (y_{j-1}-y_{j})^2}$ is the distance between a spline point and the next one. The total time it takes for a gas parcel at point $j$ to reach the nuclear ring is then
\begin{equation} \label{eq:taccr}
    t_{{\rm accr}, j} =\sum_{k=1}^{j} \di t_{{\rm accr}, k}
\end{equation}

\subsubsection{Mass}
\label{subsec:mass}

We assign a total gas mass to each point along the spline as follows (see for example \citealt{Rosolowsky2021}). First, we calculate the total CO luminosity associated to point $j$ on the spline as:
\begin{equation} \label{eq:alphaCO}
   \di L_{{\rm CO},j} = A_{\rm pix}  \Delta v \sum_k T_k  \,,
\end{equation}
where $A_{\rm pix}$ is the projected physical area of a cube pixel in $\pc^2$, $\Delta v$ is the channel width in $\kms$, $T_k$ is the brightness temperature of the cube pixel $k$ measured in K, and the sum over $k$ is extended over all pixels in the datacube that are closer to the spline point $j$ than to any other spline point (and that are associated to the bar lane according to the manual identification). The resulting $L_{\rm CO}$ has units of $\mathrm{K} \kms \pc^2$.

\newcommand{\CO}[2]{\mbox{$\mathrm{CO}\,(#1\text{--}#2)$}}
To convert this CO line luminosity into molecular gas mass, we use a \CO21-to-H$_2$ conversion factor, $\alpha_{{\rm CO}}^{(2\mhyphen1)}$, that varies spatially as a function of metallicity and total (gas plus stellar) mass surface density, following \citet{Bolatto2013}.
We favour this location-dependent conversion factor over the constant, ``galactic'' value commonly assumed in the literature, as the former can better capture the expected decrease of $\alpha_{{\rm CO}}^{(2\mhyphen1)}$ in the centres of barred galaxies due to higher gas temperature and stronger velocity gradient \citep[e.g.,][]{Sandstrom2013,Israel2020,Teng2022}.

We calculate $\alpha_{{\rm CO}}^{(2\mhyphen1)}$ at each spatial location by combining the $\alpha_{{\rm CO}}^{(1\mhyphen0)}$ prescription in \citet[Eq.~31\footnote{Note that a fixed molecular cloud surface density of $100\rm\,M_\odot\,pc^{-2}$ is adopted for this calculation to avoid unphysical conversion factor values (see discussions in \citealt{Sun2023}).}]{Bolatto2013} with a fixed \CO21/(1--0) ratio of $R_{21}=0.65$
\citep{denBrok2021,Yajima2021,Leroy2022}.
The metallicity at each location is determined from the reported radial metallicity gradient of NGC~1097 in \citet{Pilyugin2014}.
The (kpc-scale) stellar mass surface density is derived from archival Spitzer IRAC 3.6~\micron\ images, assuming an IR color-dependent stellar mass-to-light ratio following \citet{Leroy2021survey}.
Combining the location-dependent metallicity, stellar mass surface density, and kpc-scale \CO21\ line intensity, we iteratively solve for $\alpha_{{\rm CO}}^{(2\mhyphen1)}$ until its value is consistent with Eq.~31 in \citet{Bolatto2013} given the sum of kpc-scale stellar and molecular gas mass surface densities, the latter of which depends on $\alpha_{{\rm CO}}^{(2\mhyphen1)}$ itself \citep[also see Appendix~B in][]{Sun2022}.

We then convert the \CO21\ line luminosity in each pixel into a molecular gas mass via
\begin{equation} \label{eq:Mj}
    \di M_j = \alpha_{{\rm CO},j}^{(2\mhyphen1)} L_{{\rm CO},j}\,,
\end{equation}
where $\alpha_{{\rm CO},j}^{(2\mhyphen1)}$ is the \CO21-to-H$_2$ conversion factor at point $j$ (see Fig.~\ref{fig:alphaCO}).
This gives us a total mass associated with each of the $N$ points along the spline.

\subsubsection{Inflow rate}

The final step is to calculate the inflow rate. This is given as
\begin{equation} \label{eq:Mdot}
    \dot{M}_j = \epsilon \frac{\di M_j}{\di t_{{\rm accr},j}} 
\end{equation}
where $\epsilon$ is an efficiency factor that takes into account that not all the gas falling along the bar lanes is accreted as soon as it reaches the nuclear ring. Indeed, as described in the introduction, once the gas reaches the nuclear ring it partially accretes onto it, and partially ``overshoots'', to be accreted at a later stage. 

Estimates of the efficiency factor $\epsilon$ are scarce in the literature. \cite{Regan1997} estimated the efficiency factor in NGC 1530 to be $\sim 20\%$ using hydrodynamical simulations, but unfortunately this determination is inaccurate due to an inherited sign mistake in the treatment of the bar potential in the CMHOG code they used \citep{Kim2012}. \cite{Kim2011} found $\epsilon \sim 15\%$ using SPH simulations that, by today's standard, are relatively low resolution ($500\Msun/{\rm particle}$). \citet{Hatchfield2021} used simulations with much higher resolution ($25 \Msun/{\rm cell}$) and estimated the efficiency factor using three different analysis methods (see their section~3). They found that all the methods agree with each other well within the uncertainties and, for the particular barred gravitational potential they used, yield an averaged efficiency factor of $\epsilon = (30 \pm 12)\%$. They also showed that this efficiency has large fluctuations on small ($\sim$1 Myr) timescales which are averaged out over longer ($\sim$30 Myr) timescales. In this paper we adopt $\epsilon = (30 \pm 12)\%$ as our fiducial value, and discuss in Section~\ref{sec:error:epsilon} potential limitations.

\begin{figure}
	\includegraphics[width=\columnwidth]{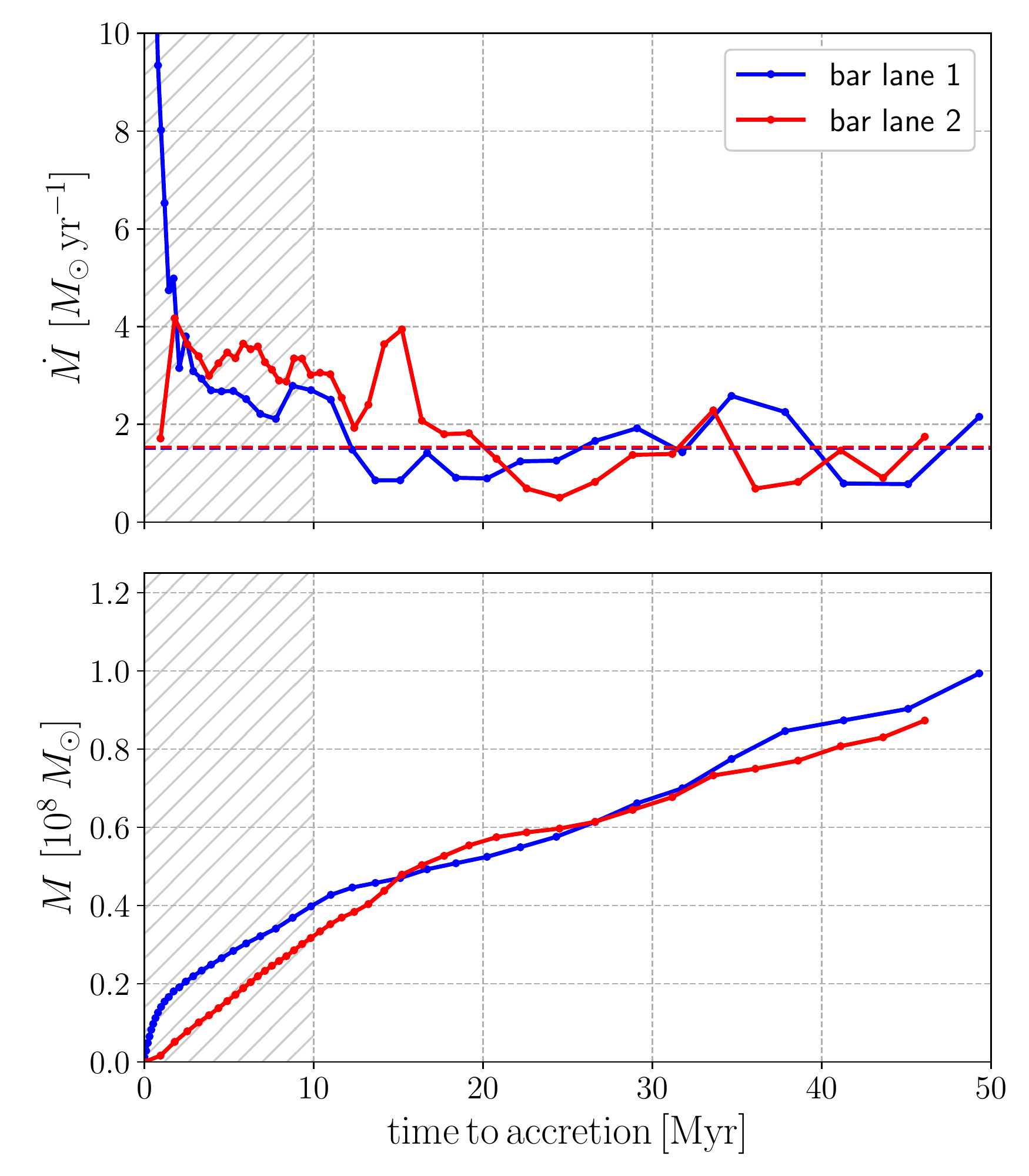}
    \caption{\emph{Top}: the instantaneous mass inflow rate along the bar lanes of NGC~1097 calculated using Eq.~\eqref{eq:Mdot}. The blue and red dashed lines show the inflow rates averaged over the non-hatched region. \emph{Bottom}: the cumulative mass accreted over time. The hatched area shows where the inflow rate is not reliable due to interaction between the bar lanes and the nuclear ring.}
    \label{fig:Mdot}
\end{figure}

\section{The bar-driven inflow rate of NGC~1097} \label{sec:results}

The top panel in Fig.~\ref{fig:vpar} shows $v_\parallel$ for the bar lanes of NGC~1097 derived using Eq.~\eqref{eq:vpar} as a function of distance from the nuclear ring, while the bottom panel shows the time to accretion calculated using Eq.~\eqref{eq:taccr}. The sharp increase in $v_\parallel$ near the centre for bar lane 1 (blue) is an artefact due to the fact that the spline becomes almost perpendicular to the line of sight when entering the ring (see Fig.~\ref{fig:dl}), causing the term $\sin\theta\to0$ in Eq.~\eqref{eq:vpar} (see discussion in Section~\ref{sec:limitations}).

The inflow rate for NGC~1097 (Eq.~\ref{eq:Mdot}), and the cumulative mass accreted over time, are shown in Fig.~\ref{fig:Mdot}. The hatched area at $t<10 \Myr$ indicates the region where we consider our inflow rate to be unreliable because the bar lanes are interacting with the nuclear ring, and the distinction between bar lanes and ring becomes blurred. As can be seen in the bottom panel of Fig.~\ref{eq:taccr}, this ``unreliable" region extends out to a distance of $d=1.7\mhyphen1.8\kpc$ from the ring. The rapid increase in the inflow rate near $t=0$ (within the unreliable region) for bar lane 1 (blue) in the top panel of Fig.~\ref{fig:Mdot} is a consequence of the spurious sharp increase in $v_\parallel$ mentioned above. Hereafter, we consider only the region between $t=10\mhyphen 50 \Myr$ for analysis.

The inflow rates on each of the two bar lanes averaged over the timespan $t=10\mhyphen50 \Myr$ are very similar, $\dot{M}_1\simeq 1.51 \Msunyr$ and $\dot{M}_2\simeq 1.53 \Msunyr$ respectively. They sum up to a total time-averaged inflow rate of $\dot{M} \simeq 3 \Msunyr$ (omitting insignificant digits). The rates are roughly constant as a function of time in the range $t=10\mhyphen40\Myr$, with fluctuations by a factor of a few over typical timescales of $\Delta t \sim 10 \Myr$ (our temporal resolution is of the order of $\sim 1 \Myr$). Peaks in the inflow rate generally correspond to denser CO clumps along the lanes visible in Fig.~\ref{fig:dl}. The $\dot{M}$ curves for each of the two lanes do not show any obvious overall upward or downward monotonic trend in the region $t=10-50\Myr$, as one would expect from a roughly steady-state flow. It is interesting to note that this occurs because the mass density along the lanes remains roughly constant as a result of two competing effects: the CO brightness increases as we move closer to the nuclear ring (Fig.~\ref{fig:dl}), while the $\alpha_{{\rm CO},j}^{(2\mhyphen1)}$ decreases (Fig.~\ref{fig:alphaCO}), so that the product of the two is roughly constant (see Eq.~\ref{eq:Mj}).

\section{Discussion} \label{sec:discussion}

\subsection{Uncertainty in the inflow rate} \label{sec:error}

\subsubsection{Uncertainty from the CO-to-H$_2$ conversion factor} \label{sec:error:alphaCO}

The largest uncertainty in our inflow rates comes from the $\alpha_{\rm CO}$ mass conversion factor (Eq.~\ref{eq:Mj} and Fig.~\ref{fig:alphaCO}). 
While we attempt to account for its spatial variation with the prescription suggested by \citet{Bolatto2013}, there are uncertainties associated with our assumption on $R_{21}$ as well as the prescription itself (see Section~\ref{subsec:mass}).
Specifically, we adopt a fixed $R_{21}=0.65$ to be consistent with the value assumed by \citet{Sandstrom2013}, which was one of the key datasets against which the \citet{Bolatto2013} prescription was calibrated.
However, $R_{21}$ is known to increase toward galaxy centres in reality, with values reaching up to 0.8--1.0 \citep{Israel2020,denBrok2021,Yajima2021,Leroy2022,Teng2022}.
While our adopted fixed $R_{21}$ ensures methodological consistency with \citet{Bolatto2013}, it might still introduce systematic uncertainties of up to 50\%.

A perhaps more important source of uncertainty comes from the \citet{Bolatto2013} prescription itself, which was calibrated against kpc-scale observations and thus should not be expected to account for small-scale effects such as variations in local gas conditions within and among individual molecular clouds.
This is demonstrated by recent studies on barred galaxy centres \citep[e.g.,][]{Teng2022,Teng2023}, which reported significant variations in optical depth and temperature at $\sim$100~pc scales, leading to 2--3 times lower $\alpha_{\rm CO}$ than the kpc-scale prediction from \citet{Bolatto2013} in bar lanes.
Therefore, using the kpc-scale stellar and gas mass surface densities consistent with  \citet{Bolatto2013}, we expect our $\alpha_{\rm CO}$ calculation to only reflect kpc-scale variations and leave systematic uncertainties at a factor of 2--3 level when applied to $\sim$100~pc scale CO data.
Considering both sources of uncertainties discussed above, we expect the $\alpha_{\rm CO}^{(2\mhyphen1)}$ factor to be uncertain by at least a factor of two, which translates into an uncertainty of $\delta\dot{M}=\pm 1.5\Msunyr$ on the inflow rate.

\subsubsection{Uncertainty from the efficiency factor} \label{sec:error:epsilon}

To calculate the error associated with the efficiency factor, we use the recommended value of $\delta \epsilon = 12\%$ from \cite{Hatchfield2021}, which translates into an error on the inflow rate of $\delta\dot{M}=\pm1.2\Msunyr$. This is a relatively generous uncertainty that takes into account potential variations in
the morphology of the bar lanes, different orbital initial conditions on the gas on them, and spatial variations of the efficiency factor along the bar lanes.

It is worth mentioning that the gravitational potential that \cite{Hatchfield2021} used to derive the efficiency factor is tuned to reproduce the properties of the Milky Way. While this can be considered a reasonably general strongly barred potential, it is possible that the efficiency factor depends on the characteristics of the bar potential (e.g.\, stronger bars might be expected to have larger efficiency factors). This topic will require further investigation using simulations that is outside of the scope of this paper.

Another element that we ignore in this paper but could affect the efficiency factor is magnetic fields. Magnetic stresses can help removing further angular momentum from the infalling gas, thus leading to increased inflow rates \citep{Kim2012b}. Indeed, magnetic stresses have been suggested to be an important ingredient in the fuelling of the nuclear ring of NGC~1097 \citep{Beck2005,Lopez-Rodriguez2021}. Thus, it might turn out that the efficiency factor should be increased when magnetic fields are taken into account.

Finally, we have neglected that a small amount of star formation might occur while the gas is traversing the bar lanes. However, this will give a contribution of few percent at most since the dynamical timescale that it takes for the gas to cross the lanes ($\sim 40 \Myr$) is small compared to typical gas star formation depletion times ($\sim 1 \Gyr$).

\subsubsection{Uncertainty from the bar pattern speed and galaxy orientation} \label{sec:error:other}

\begin{figure}
	\includegraphics[width=\columnwidth]{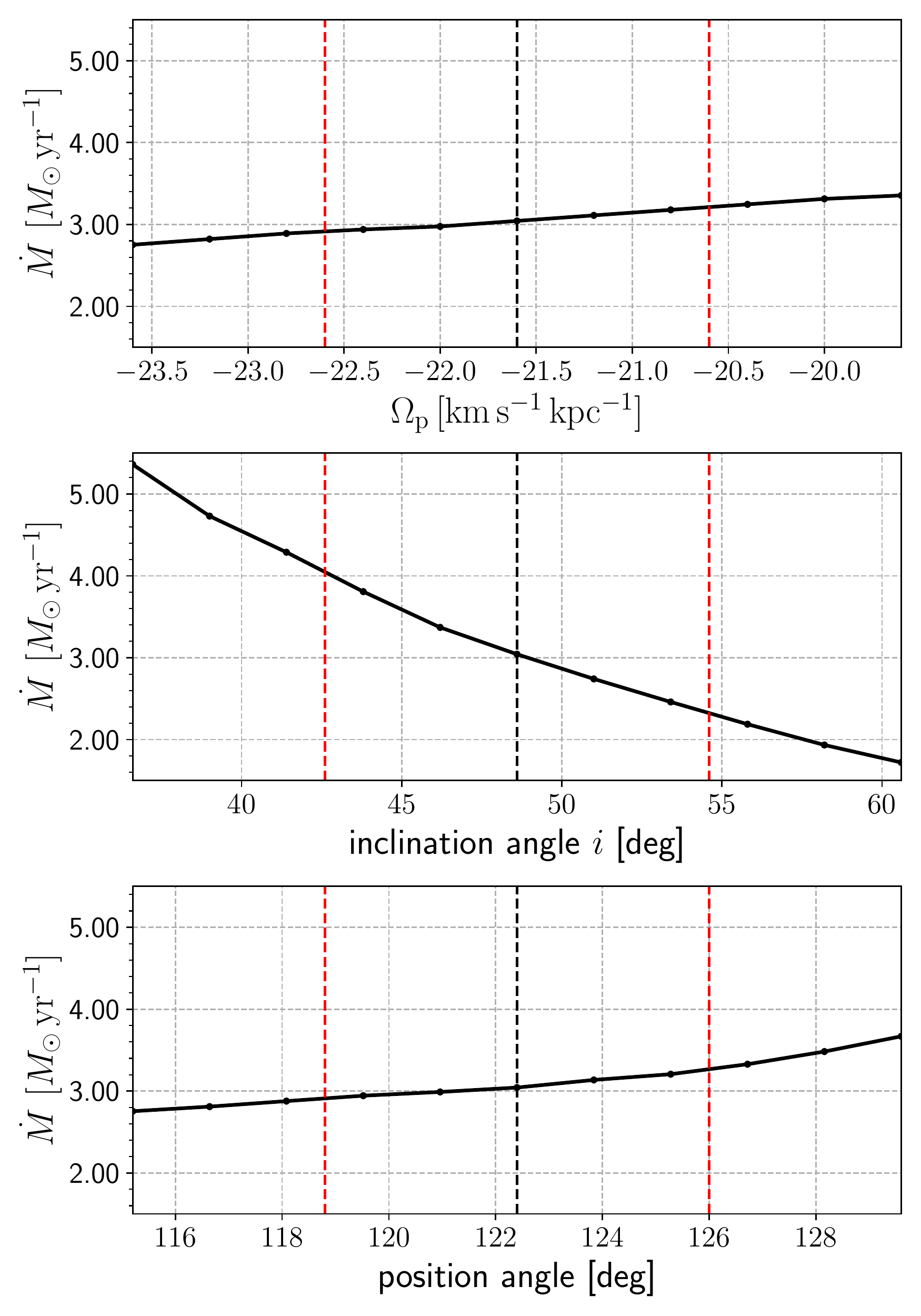}
    \caption{Time-averaged total inflow rate in NGC~1097 as a function of assumed parameters. \emph{Top}: bar pattern speed. \emph{Middle}: inclination angle. \emph{Bottom}: position angle. The black dashed lines indicate the fiducial values. The red dashed lines indicate the 1-$\sigma$ uncertainty ranges.}
    \label{fig:errors}
\end{figure}

Further sources of uncertainty are the assumed inclination angle $i$, position angle, and bar pattern speed $\Omegap$. Figure~\ref{fig:errors} shows how the total time-averaged inflow rate changes if these parameters are varied between plausible ranges, where the red dashed vertical lines represent the 1-$\sigma$ uncertainties. The uncertainty in the bar pattern speed is based on \cite{Lin2013}, while the uncertainties in the inclination and position angles are taken from table~2 of \cite{Lang2020}. Table~\ref{tab:error} shows the corresponding errors on the inflow rates.

\subsubsection{Summary}

Summing all sources of errors in quadrature, our estimate for the total inflow rate obtained summing both bar lanes in NGC~1097 is $\dot{M} = 3.0 \pm 2.1 \Msunyr $. Table~\ref{tab:error} provides a summary of all the sources of uncertainty discussed above and their associated error.

    \begin{table}
    	\centering
    	\begin{tabular}{lc} 
        Source of error & Contribution to $\delta\dot{M}$ \\
    		\hline
      $\alpha_{\rm CO}$ & $ 1.5 \Msunyr$ \\
      $\epsilon$ & $ 1.2\Msunyr$ \\
      $\Omegap$ & $ 0.15 \Msunyr$ \\
      inclination angle $i$ & $ 0.9 \Msunyr$ \\
      position angle & $ 0.2 \Msunyr$ \\
      \hline
      \hline
      Total in quadrature & $ 2.1 \Msunyr$
    \end{tabular}
            \caption{Summary of uncertainties on the total inflow rate.}
    \label{tab:error}
    \end{table}

\subsection{Mass balance in the nuclear ring} \label{sec:balance}

Mass conservation in a cylindrical volume containing the nuclear ring (e.g.\ $R< 1 \, \kpc$ and $|z|<300\, \pc$) implies:
\begin{equation}
    \dot{M} = {\rm SFR} +  \dot{M}_{\rm out} + \dot{M}_{\rm ring},
\end{equation}
where $\dot{M}$ is the bar-driven accretion rate onto the ring, SFR is the star formation rate in the ring, $\dot{M}_{\rm out}$ is the mass lost from the ring due to outflowing gas, and $\dot{M}_{\rm ring}$ is the rate of change of the total gas mass in the ring. In this paper we have found $\dot{M} = 3.0 \pm 2.1 \Msunyr$. There is tentative evidence for a central molecular outflow in NGC~1097, with an estimated rate of $\dot{M}_{\rm out} \sim 0.6, \Msunyr$, but further analysis are required to confirm this \citep{Stuber2021}. 

The star formation rate in the nuclear ring has been estimated to be ${\rm SFR}\simeq 1.8 \mhyphen 2 \Msunyr$ \citep{Sandstrom2010,Hsieh2011,Prieto2019,Lopez-Rodriguez2021,Song2021}. Thus, most of the currently inflowing gas is being consumed by star formation. This suggests that the inflow is causally controlling the SFR in the nuclear ring of NGC~1097, as predicted by the hydrodynamical simulations of \cite{Seo2019}, \cite{Sormani2020b} and \cite{Moon2021b,Moon2021a}.

For comparison, in the Milky Way we currently have $\dot{M}\simeq 0.8 \pm 0.6 \Msunyr$, $\dot{M}_{\rm out} \gtrsim 0.6 \Msunyr$ and ${\rm SFR}\simeq0.1\Msunyr$ \citep{Henshaw2022}. Thus, both NGC~1097 and the Milky Way appear compatible with a rough mass balance between inflow, outflow and star formation rate, without a net mass accumulation in the ring ($\dot{M}_{\rm ring}\simeq0$) that is sometimes invoked to eventually lead to a future burst of star formation.

\subsection{Limitations of Equation~\eqref{eq:vpar}} \label{sec:limitations}

Equation~\eqref{eq:vpar} has $\sin(\theta)$ in the denominator. This term goes to zero when the bar lane becomes perpendicular to the line of sight. Thus, our method produces large errors when a galaxy has this orientation. For NGC~1097 this is not a significant issue, since it happens only in the innermost part of bar lane 1, which, as discussed in Section~\ref{sec:results}, does not affect our estimation of the inflow rates. For galaxies where the bar lanes are almost parallel to the line of nodes (e.g. NGC 1300, \citealt{Lang2020}) this issue becomes more important and some other way to estimate $v_\parallel$ should be used.

Another limitation of Eq.~\eqref{eq:vpar} is the following. This equation makes sense only if $v_\parallel>0$, i.e.\ the numerator and denominator have the same sign, otherwise the gas would be climbing upstream the bar lanes, which is unphysical within our simplified steady-state picture. In practice, this condition means that the $\Omegap x \sin i$ line (dashed line in the bottom panels of Fig.~\ref{fig:dl}) and $\vlos$ from the spline interpolation (blue and red line in the same figure) should never cross. This condition is satisfied for NGC~1097. However, it might not be satisfied in other galaxies. Numerical simulations from \citet{Tress2020a} show that it can happen that some portions of the bar lane move upstream for a brief moment of time, i.e. ``against the current'' (see their figure 16).  As explained in their section 5, this happens when gas on the bar lane collides with clouds overshooting from the bar lane on the other side. We found preliminary evidence of a similar behaviour  in NGC 4535 when visually inspecting the PHANGS-ALMA CO datacube. The dynamics of these events are not taken into account in our simplified formula based on a steady-state picture.

\subsection{Can we trust the fluctuations in inflow rate?}

As mentioned in Sect.~\ref{sec:results}, the inflow rates shown in Figure~\ref{fig:Mdot} at $t\geq 10 \Myr$ fluctuate by a factor of a few as a function of time over typical timescales $\Delta t \sim 10 \Myr$. Will these fluctuations correspond to real temporary boosts in the future inflow rate?

Closer inspection reveals, as one might expect, that the peaks in the top panel of Fig.~\ref{fig:Mdot} correspond to dense clumps/clouds in the CO distribution in Fig.~\ref{fig:dl}. When these clumps reach the nuclear ring, they will likely cause a boost in the inflow rate. However, due to the stochasticity of the accretion process, it might happen that one of these clouds overshoots and misses the nuclear ring entirely \citep[e.g.][]{Hatchfield2021}. Thus, the more precise answer to the question above is the following: while Fig.~\ref{fig:Mdot} gives a fairly good idea of the order of magnitude of the typical fluctuations, it can predict when the inflow rates will increase only in a probabilistic sense.

\subsection{Comparison to previous works}

\cite{Prieto2019} crudely estimated the mass rate arriving at the ring by approximating the bar lane as a cylinder with diameter $D=600\pc$ and using the formula $\dot{M}=2 D^2 n ff_{\rm lane} v_{\rm flow}$, where $ff_{\rm lane}=0.02$ is their estimated ``area'' filling factor, $n$ is their estimated volume density, and $v_{\rm flow}=350\kms$ is their estimated flow velocity at the entrance of the ring. They obtained $\dot{M}\gtrsim 3\Msunyr$, consistent with our estimate. The fact that the two estimates are so similar is likely a coincidence due to fortuitous balancing of factors in their formula given the fact that the individual factors are quite different from the values used in this paper (e.g.\ their flow velocity is much higher than the velocities we estimated in Fig.~\ref{fig:vpar}), their large uncertainties in the individual factors, and that their estimate only takes into account the ``arrival" region very close to the ring, which we have excluded from our analysis (see Section~\ref{sec:results}).

\section{Conclusion} \label{sec:conclusion}

We have applied a simple geometrical method to PHANGS-ALMA CO data to derive the mass inflow rate onto the nuclear ring of the barred galaxy NGC~1097. The method allows one to estimate the inflow rate given the mass density and line-of-sight velocity along the bar bar lanes, the bar pattern speed and the galaxy orientation. The PHANGS-ALMA CO data set is ideal for this purpose thanks to having the high-spatial resolution needed to resolve the bar lanes, as well as the sensitivity and the short-spacing data to recover a full picture of the emission, both of which are required for accurate mass measurements. 

We found a total inflow rate of $\dot{M}=(3.0 \pm 
2.1)\Msunyr$ averaged over a timespan of 40 Myr which fluctuates by a factor of few over a typical timescale $\Delta t \sim 10 \Myr$. The main uncertainty on our estimated value of the inflow rate comes from the $\alpha_{{\rm CO},j}^{(2\mhyphen1)}$ factor used to convert $^{12}$CO $J=2\to 1$ line luminosity to gas mass. 

Most of the inflow is currently being consumed by star formation in the nuclear ring at a rate ${\rm SFR} \simeq 1.8 \mhyphen 2 \Msunyr$ \citep{Hsieh2011,Prieto2019,Song2021}. This suggests that the inflow is causally controlling the star formation in the ring, as predicted by simulations of star formation in nuclear rings of barred galaxies \citep{Seo2019,Sormani2020b,Moon2021b,Moon2021a}.

\section*{Acknowledgements}

MCS and RSK thank for support from the European Research Council via the ERC Synergy Grant ``ECOGAL – Understanding our Galactic ecosystem: from the disk of the Milky Way to the formation sites of stars and planets'' (grant 855130),  from  the Heidelberg Cluster of Excellence (EXC 2181 - 390900948) ``STRUCTURES'', funded by the German Excellence Strategy, and from the German Ministry for Economic Affairs and Climate Action in project ``MAINN'' (funding ID 50OO2206). They also acknowledge  computing resources provided by the Ministry of Science, Research and the Arts (MWK) of the State of Baden-W\"{u}rttemberg through bwHPC and DFG through grant INST 35/1134-1 FUGG and for data storage at SDS@hd through grant INST 35/1314-1 FUGG. MCS acknowledges financial support from the Royal Society (URF\textbackslash R1\textbackslash 221118).
ATB and FB acknowledge funding from the European Research Council (ERC) under the European Union’s Horizon 2020 research and innovation programme (grant agreement No.726384/Empire).
The work of JS is partially supported by the Natural Sciences and Engineering Research Council of Canada (NSERC) through the Canadian Institute for Theoretical Astrophysics (CITA) National Fellowship. JMDK acknowledges funding from the European Research Council (ERC) under the European Union's Horizon 2020 research and innovation programme via the ERC Starting Grant MUSTANG (grant agreement number 714907). COOL Research DAO is a Decentralised Autonomous Organisation supporting research in astrophysics aimed at uncovering our cosmic origins. J.\ D.\ Henshaw gratefully acknowledges financial support from the Royal Society (University Research Fellowship; URF\textbackslash R1\textbackslash 221620). RJS gratefully acknowledges an STFC Ernest Rutherford fellowship (grant ST/N00485X/1). RCL acknowledges support provided by a National Science Foundation (NSF) Astronomy and Astrophysics Postdoctoral Fellowship under award AST-2102625. MQ acknowledges support from the Spanish grant PID2019-106027GA-C44, funded by MCIN/AEI/10.13039/501100011033. ES and JN acknowledge funding from the European Research Council (ERC) under the European Union’s Horizon 2020 research and innovation programme (grant agreement No. 694343).
KG is supported by the Australian Research Council through the Discovery Early Career Researcher Award (DECRA) Fellowship (project number DE220100766) funded by the Australian Government. 
KG is supported by the Australian Research Council Centre of Excellence for All Sky Astrophysics in 3 Dimensions (ASTRO~3D), through project number CE170100013.
SKS acknowledges financial support from the German Research Foundation (DFG) via Sino-German research grant SCHI 536/11-1. Y.-H.T. acknowledges funding support from NRAO Student Observing Support Grant SOSPADA-012 and from the National Science Foundation (NSF) under grant No. 2108081. EJW gratefully acknowledges funding from the German Research Foundation (DFG) in the form of an Emmy Noether Research Group (grant number KR4598/2-1, PI Kreckel).
CE acknowledges funding from the Deutsche Forschungsgemeinschaft (DFG) Sachbeihilfe, grant number BI1546/3-1.

%%%%%%%%%%%%%%%%%%%%%%%%%%%%%%%%%%%%%%%%%%%%%%%%%%
\section*{Data Availability}

The data underlying this article are from the ALMA project 
ADS/JAO.ALMA\#2017.1.00886.L, and are available through the following link: \url{https://www.canfar.net/storage/list/phangs/RELEASES/PHANGS-ALMA/}. ALMA is a partnership of ESO (representing its member states), NSF (USA) and NINS (Japan), together with NRC (Canada), MOST and ASIAA (Taiwan), and KASI (Republic of Korea), in cooperation with the Republic of Chile. The Joint ALMA Observatory is operated by ESO, AUI/NRAO and NAOJ.

%%%%%%%%%%%%%%%%%%%% REFERENCES %%%%%%%%%%%%%%%%%%

\bibliographystyle{mnras}
\bibliography{bibliography}

%%%%%%%%%%%%%%%%%%%%%%%%%%%%%%%%%%%%%%%%%%%%%%%%%%

%%%%%%%%%%%%%%%%% APPENDICES %%%%%%%%%%%%%%%%%%%%%

% \appendix

%%%%%%%%%%%%%%%%%%%%%%%%%%%%%%%%%%%%%%%%%%%%%%%%%%

% Don't change these lines
\bsp	% typesetting comment
\label{lastpage}
\end{document}